\documentclass[fleqn,twoside]{article}
\usepackage{espcrc2}

\usepackage{graphicx}
\usepackage[figuresright]{rotating}

\newcommand{\AmS}{{\protect\the\textfont2
  A\kern-.1667em\lower.5ex\hbox{M}\kern-.125emS}}

\def\lcdm{$\Lambda$CDM}
\def\gsim {\lower .1ex\hbox{\rlap{\raise .6ex\hbox{\hskip .3ex
        {\ifmmode{\scriptscriptstyle >}\else
                {$\scriptscriptstyle >$}\fi}}}
        \kern -.4ex{\ifmmode{\scriptscriptstyle \sim}\else
                {$\scriptscriptstyle\sim$}\fi}}}
\def\lsim {\lower .1ex\hbox{\rlap{\raise .6ex\hbox{\hskip .3ex
        {\ifmmode{\scriptscriptstyle <}\else
                {$\scriptscriptstyle <$}\fi}}}
        \kern -.4ex{\ifmmode{\scriptscriptstyle \sim}\else
                {$\scriptscriptstyle\sim$}\fi}}}
\def\kms{{\rm km}\, {\rm s}^{-1}}
\def\kmsMpc{{\rm km}\, {\rm s}^{-1}\,{\rm Mpc}^{-1}}
\def\hMpc{h^{-1}{\rm Mpc}}

\def\cvir{c_{\rm vir}}

\def\rs{r_{\rm s}}
\def\Rvir{R_{\rm vir}}

\def\Vmax{V_{\rm max}}

\def\lcdm{$\Lambda$CDM}

\title{Status of Cold Dark Matter Cosmology}
       
\author{Joel R. Primack\address{Physics Department, University of California,
        Santa Cruz, CA 95064 USA}}
\begin{document}

\begin{abstract}
Cold Dark Matter (CDM) has become the standard modern theory of
cosmological structure formation.  Its predictions appear to be in
good agreement with data on large scales, and it naturally accounts
for many properties of galaxies.  But despite its many successes,
there has been concern about CDM on small scales because of the
possible contradiction between the linearly rising rotation curves
observed in some dark-matter-dominated galaxies vs.~the $1/r$ density
cusps at the centers of simulated CDM halos.  Other CDM issues on
small scales include the very large number of small satellite halos in
simulations, far more than the number of small galaxies observed
locally, and problems concerning the angular momentum of the baryons
in dark matter halos.  The latest data and simulations have lessened,
although not entirely resolved, these concerns.  Meanwhile, the main
alternatives to CDM that have been considered to solve these problems,
self-interacting dark matter (SIDM) and warm dark matter (WDM), have
been found to have serious drawbacks.
\vspace{1pc}
\end{abstract}

% typeset front matter (including abstract)
\maketitle

\section{Introduction}

The universe on the largest scales can be described by three numbers:
\begin{itemize}
 \item $H_0 \equiv 100 h \kmsMpc$, the Hubble parameter (expansion
rate of the universe) at the present epoch, 
 \item $\Omega_m \equiv \rho/\rho_c$, the density of matter $\rho$ in 
units of critical density $\rho_c \equiv 3 H_0^2 (8\pi G)^{-1} = 2.78 
\times 10^{11} h^2 M_\odot$ Mpc$^{-3}$, and
 \item $\Omega_\Lambda \equiv \Lambda (3H_0^2)^{-1}$, the
corresponding quantity for the cosmological constant.
\end{itemize}
The current values of these and other key parameters are
summarized in the Table below (for additional references and
discussion see \cite{isss}).  It remains to be seen whether the
``dark energy'' represented by the cosmological constant $\Lambda$ is
really constant, or is perhaps instead a consequence of the dynamics of
some fundamental field as in ``quintessence''
theories \cite{quintessence}.

Cold Dark Matter (CDM) assumes that the dark matter is mostly cold ---
i.e., with negligible thermal velocities in the early universe, either
because the dark matter particles are weakly interacting massive
particles (WIMPs) with mass $\sim 10^2$ GeV, or alternatively because
they are produced without a thermal distribution of velocities, as is
the case with axions.  CDM also assumes that the fluctuations in the
dark matter are adiabatic and have a nearly Zel'dovich spectrum.
Considering that the CDM model of structure formation in the universe
was proposed almost twenty years ago \cite{peeb82,bfpr}, its successes
are nothing short of amazing.  As I will discuss, the \lcdm\ variant
of CDM with $\Omega_m = 1-\Omega_\Lambda \approx 0.3$ appears to be in
good agreement with the available data on large scales.  Issues that
have arisen on smaller scales, such as the centers of dark matter
halos and the numbers of small satellites, have prompted people to
propose a wide variety of alternatives to CDM, such as warm dark
matter (WDM) \cite{bode} and self-interacting dark matter (SIDM) \cite{sidm}. It
remains to be seen whether such alternative theories with extra
parameters actually turn out to be in better agreement with data.  As
I will discuss below, it now appears that WDM and SIDM are both
probably ruled out, while the small-scale predictions of CDM may be in
better agreement with the latest data than appeared to be the case as
recently as a year ago.

In the next section I will briefly review the current observations
and the successes of \lcdm\ on large scales, and then I will
discuss the possible problems on small scales.

\section{Cosmological Parameters and Observations on Large Scales}

The table below\footnote{Updating the one in my talk at DM2000
\protect\cite{dm2000}.} summarizes the current observational
information about the cosmological parameters, with estimated
$1\sigma$ errors.  The quantities in brackets have been deduced using
at least some of the \lcdm\ assumptions.  Is is apparent that there is
impressive agreement between the values of the parameters determined
by various methods, including those based on \lcdm.  In particular,
(A) several different approaches all suggest that $\Omega_m \approx
0.3$; (B) the location of the first acoustic peak in the CMB angular
anisotropy power spectrum, now very well determined independently by
the BOOMERANG \cite{lange} and MAXIMA1 \cite{maxima} balloon data 
and by the DASI interferometer at the South Pole
\cite{pryke}, implies that $\Omega_m+\Omega_\Lambda \approx 1$; and
(C) the data on supernovae of Type Ia (SNIa) at redshifts $z=0.4-1.2$
from two independent groups imply that $\Omega_\Lambda - {4\over 3}
\Omega \approx {1\over 3}$.  Any two of these three results then imply
that $\Omega_\Lambda \approx 0.7$.  The $1\sigma$ errors in these
determinations are about 0.1.

Questions have been raised about the reliability of the high-redshift
SNIa results, especially the possibilities that the SNIa properties at
high redshift might not be sufficiently similar to those nearby to use
them as standard candles, and that there might be ``grey'' dust (which
would make the SNIa dimmer but not change their colors).  Although the
available evidence disfavors these possibilities,\footnote{For
example, SNIa at $z=1.2$ and $\sim1.7$ apparently have the brightness
expected in a \lcdm\ cosmology but are brighter than would be expected
with grey dust, and the infrared brightness of a nearer SNIa is also
inconsistent with grey dust \cite{SNIa_IR}.}  additional observations
are needed on SNIa at high redshift, both to control systematic
effects and to see whether the dark energy is just a cosmological
constant or is perhaps instead changing with redshift as expected in
``quintessence'' models \cite{quintessence}.  But it is important to
appreciate that, independently of (C) SNIa, (A) cluster and other
evidence for $\Omega_m \approx 0.3$ \cite{Turner}, together with (B)
$\sim1^\circ$ CMB evidence for $\Omega_m + \Omega_\Lambda \approx 1$,
imply that $\Omega_\Lambda \approx 0.7$.

\begin{table}
\caption{Cosmological Parameters [results assuming \lcdm\ in
brackets]}
\label{ta:parameters}
%\centerline{\vbox{\halign{\ \ #\hfill \quad \qquad &$#$\hfill \ &$#$
%&#\hfill \ \ \cr
\centerline{\vbox{\halign{\ $#$\hfill \ &$#$ &#\hfill \ \ \cr
\noalign{\hrule}
\noalign{\vskip .10in}
H_0          &= &$100 \,h$ km s$^{-1}$ Mpc$^{-1}$ ,
                                \ $h = 0.7 \pm 0.08$ \cr
t_0          &= &$13\pm2$ Gyr (from globular clusters) \cr
{}           &= &$[14\pm0.5$ Gyr, \lcdm\ + CMB] \cr
\Omega_b     &= &$(0.040\pm0.002)h_{70}^{-2}$ (from D/H) \cr
{}           &> &[$0.035 h_{70}^{-2}$ from Ly$\alpha$ forest opacity] \cr
\Omega_m     &= &$0.33\pm0.035$ (from cluster baryons etc.) \cr
{}           &= & [$0.34\pm0.1$ from Ly$\alpha$ forest $P(k)$] \cr
{}           &= & [$0.4\pm0.2$ from cluster evolution] \cr
%{}           &> & $0.3$ ($2.4 \sigma$, from cosmic flows) \cr
{}           &\approx &$\frac{3}{4} \Omega_\Lambda - \frac{1}{4}
                        \pm{\frac{1}{8}}$ from SN Ia \cr
\Omega_{tot} &= &$1.04\pm0.05$ (from CMB peak location) \cr
\Omega_\Lambda &= &$0.73 \pm 0.08$ (from previous two lines) \cr
{}           &< & 0.73 (2$\sigma$) from radio QSO lensing \cr
\Omega_\nu  &$\gsim$ &0.001 (from SuperKamiokande data) \cr
{}           &$\lsim$ &[0.05 in \lcdm-type models] \cr
\noalign{\vskip .10in}
\noalign{\hrule}
}}}
\end{table}

All methods for determining the Hubble parameter now give compatible
results, confirming our confidence that this crucial parameter has now
been measured robustly to a $1\sigma$ accuracy of about $10\%$.  The
final result\cite{freedman} from the Hubble Key Project on the
Extragalactic Distance Scale is $72\pm8 \kmsMpc$, or $h=0.72\pm0.08$,
where the stated error is dominated by one systematic uncertainty, the
distance to the Large Magellanic Cloud (used to calibrate the Cepheid
period-luminosity relationship).  The most accurate of the direct
methods for measuring distances $d$ to distant objects, giving the
Hubble parameter directly as $H_0=d/v$ where the velocity is
determined by the redshift, are (1) time delays between luminosity
variations in different gravitationally lensed images of distant
quasars, giving $h\approx 0.65$, and (2) the Sunyaev-Zel'dovich effect
(Compton scattering of the CMB by the hot electrons in clusters of
galaxies), giving $h \approx 0.63$ \cite{Carlstrom,freedman}. For the
rest of this article, I will take $h=0.7$ whenever I need to use an
explicit value, and express results in terms of $h_{70} \equiv H_0/70
\kmsMpc$.

If $\Omega_{tot} = 1$ and structure formed from adiabatic initial
conditions as assumed in \lcdm, CMB data imply $t_0=14.0 \pm0.5$
Gyr~\cite{knox}.  For a \lcdm\ universe with $\Omega_m=(0.2) 0.3
(0.4,0.5)$, the expansion age is $t_0=(15.0) 13.47 (12.41,11.61)
h_{70}^{-1}$ Gyr.  For $\Omega_m \approx 0.3$ and
$h\approx0.7$, there is excellent agreement with the latest estimates
of the ages of the oldest stars in the Milky Way, both (A) from the
globular cluster Main Sequence turnoff luminosities \cite{carretta},
giving $12-13 \pm2$ Gyr, (B) using the thorium and uranium radioactive
decay chronometers for halo stars \cite{sneden}, giving $14\pm3$ Gyr
and $12.5\pm3$ Gyr, respectively, and (C) from white dwarf cooling
time, giving $12.7\pm0.7$ Gyr as the age of the globular cluster M4
\cite{Hansen}.  It is remarkable that these four different clocks all
agree!

The lower limit on the hot dark matter (i.e. neutrino) contribution to
the cosmological density comes from the Super-Kamiokande atmospheric
neutrino data~\cite{superk,beamline}.  The latest upper limit is from
the 2dF redshift survey galaxy power spectrum~\cite{2dfneutrino}.

\section{Further Successes of \lcdm}

The \lcdm\ cosmology correctly predicts the abundances of clusters
nearby and at $z\lsim1$ within the current uncertainties in the values
of the parameters.  It is even consistent with $P(k)$ from the
Ly$\alpha$ forest \cite{croft} and from CMB anisotropies.
Low-$\Omega_m$ CDM predicts that the amplitude of the power spectrum
$P(k)$ is rather large for $k \lsim 0.02 h/{\rm Mpc}^{-1}$, i.e. on
size scales larger ($k$ smaller) than the peak in $P(k)$.  The
largest-scale surveys, 2dF and SDSS, should be able to measure $P(k)$
on these scales and test this crucial prediction soon; preliminary
results are encouraging \cite{powerspectrum}.

The hierarchical structure formation which is inherent in CDM already
explains why most stars are in big galaxies like the Milky
Way \cite{bfpr}: smaller galaxies merge to form these larger ones,
but the gas in still larger structures takes too long to cool to
form still larger galaxies, so these larger structures --- the largest
bound systems in the universe --- become groups and clusters instead of
galaxies.  

What about the more detailed predictions of \lcdm, for example on the
spatial distribution of galaxies.  On large scales, there appears to
be a pretty good match.  In order to investigate such questions
quantitatively on the smaller scales where the best data is available
it is essential to do N-body simulations, since the mass fluctuations
$\delta\rho/\rho$ are nonlinear on the few-Mpc scales that are
relevant.  My colleagues and I were initially concerned that \lcdm\
would fail this test \cite{kph}, since the dark matter power spectrum
$P_{dm}(k)$ in \lcdm, and its Fourier transform the correlation
function $\xi_{dm}(r)$, are seriously in disagreement with the galaxy
data $P_g(k)$ and $\xi_g(r)$.  One way of describing this is to say
that scale-dependent antibiasing is required for \lcdm\ to agree with
observations.  That is, the bias parameter $b(r) \equiv
[\xi_g(r)/\xi_{dm}(r)]^{1/2}$, which is about unity on large scales,
must decrease to less than 1/2 on scales of a few
Mpc~\cite{kph,jenkins}. This was the opposite of what was expected:
galaxies were generally thought to be more correlated than the dark
matter on small scales.  However, when it became possible to do
simulations of sufficiently high resolution to identify the dark
matter halos that would host visible galaxies \cite{overcoming,colin},
it turned out that their correlation function is essentially identical
with that of observed galaxies!  

Jim Peebles, who largely initiated the study of galaxy correlations
and first showed that $\xi_g(r) \approx (r/r_0)^{-1.8}$ with $r_0
\approx 5 \hMpc$ \cite{davispeeb}, thought that this simple power law
must be telling us something fundamental about cosmology.  However, it
now appears that the power law $\xi_g$ arises because of a coincidence
-- an interplay between the non-power-law $\xi_{dm}(r)$ and the
decreasing survival probability of dark matter halos in dense regions
because of their destruction and merging.  But the essential lesson is
that \lcdm\ correctly predicts the observed $\xi_g(r)$.

The same theory also predicts the number density of galaxies.  Using
the observed correlations between galaxy luminosity and internal
velocity, known as the Tully-Fisher and Faber-Jackson relations for
spiral and elliptical galaxies respectively, it is possible to convert
observed galaxy luminosity functions into approximate galaxy velocity
functions, which describe the number of galaxies per unit volume as a
function of their internal velocity.  The velocity function of dark
matter halos is robustly predicted by N-body simulations for CDM-type
theories, but to connect it with the observed internal velocities of
bright galaxies it is necessary to correct for the infall of the
baryons in these galaxies \cite{bffp,mmw,kochanekwhite}, which must
have happened to create their bright centers and disks.  When we did
this it appeared that \lcdm\ with $\Omega_m=0.3$ predicts perhaps too
many dark halos compared with the number of observed galaxies with
internal rotation velocities $V \approx 200 \kms$
\cite{vfun,cole00}. While the latest results from the big surveys now
underway appear to be in better agreement with these \lcdm\
predictions~\cite{2dF,sdss}, this is an important issue that is being
investigated in detail~\cite{pahre}.  Questions concerning the luminosity
function still need to be resolved \cite{Wright}.

The problem just mentioned of accounting for baryonic infall is just
one example of the hydrodynamical phenomena that must be taken into
account in order to make realistic predictions of galaxy properties in
cosmological theories.  Unfortunately, the crucial processes of
especially star formation and supernova feedback are not yet well
enough understood to allow reliable calculations.  Therefore, rather
than trying to understand galaxy formation from full-scale
hydrodynamic simulations (for example \cite{Weinberg}), more progress
has been made via the simpler approach of semi-analytic modelling of
galaxy formation (initiated by White and
Frenk~\cite{whitefrenk,kauffmann93,cole94}, reviewed and extended by
Rachel Somerville and me~\cite{sp}).  The computational efficiency of
SAMs permits detailed exploration of the effects of the cosmological
parameters, as well as the parameters that control star formation and
supernova feedback.  We have shown \cite{sp} that both flat and open
CDM-type models with $\Omega_m = 0.3-0.5$ predict galaxy luminosity
functions and Tully-Fisher relations that are in good agreement with
observations.  Including the effects of (proto-)galaxy interactions at
high redshift in SAMs allows us to account for the observed properties
of high-redshift galaxies, but only for $\Omega_m \approx 0.3-0.5$
\cite{spf}.  Models with $\Omega_m=1$ and realistic power spectra
produce far too few galaxies at high redshift, essentially because of
the fluctuation growth rate argument mentioned above.

In order to tell whether \lcdm\ accounts in detail for galaxy
properties, it is essential to model the dark halos accurately.  The
Navarro-Frenk-White (NFW) \cite{nfw} density profile $\rho_{NFW}(r)
\propto r^{-1} (r+r_s)^{-2}$ is a good representation of typical dark
matter halos of galactic mass, except possibly in their very centers
(\S4).  Comparing simulations of the same halo with numbers of
particles ranging from $\sim10^3$ to $\sim10^6$, my colleagues and I
have also shown~\cite{klypin01} that $r_s$, the radius where the
log-slope is -2, can be determined accurately for halos with as few as
$\sim10^3$ particles.  Based on a study of thousands of halos at many
redshifts in an Adaptive Refinement Tree (ART) \cite{art} simulation
of the \lcdm\ cosmology, we \cite{bullock99} found that the
concentration $\cvir \equiv \Rvir/\rs$ has a log-normal distribution,
with $1\sigma$ $\Delta (\log \cvir) = 0.14$ at a given
mass~\cite{jing,risa01}.  This scatter in concentration results in a
scatter in maximum rotation velocities of $\Delta \Vmax/\Vmax = 0.12$;
thus the distribution of halo concentrations has as large an effect on
galaxy rotation curves shapes as the well-known log-normal
distribution of halo spin parameters $\lambda$. Frank van den Bosch
\cite{vdboschTF} showed, based on a semi-analytic model for galaxy
formation including the NFW profile and supernova feedback, that the
spread in $\lambda$ mainly results in movement along the Tully-Fisher
line, while the spread in concentration results in dispersion
perpendicular to the Tully-Fisher relation.  Remarkably, he found that
the dispersion in \lcdm\ halo concentrations produces a Tully-Fisher
dispersion that is consistent with the observed
one.\footnote{Actually, this was the case with the dispersion in
concentration $\Delta(\log \cvir) = 0.1$ found for relaxed halos by
Jing \protect\cite{jing}, while we \protect\cite{bullock99} found the
larger dispersion mentioned above.  However Risa Wechsler, in her
dissertation research with me~\cite{risa01}, found that the dispersion
in the concentration at fixed mass of the halos that have not had a
major merger since redshift $z=2$ (and could thus host a spiral
galaxy) is consistent with that found by Jing.  We also found that the
median and dispersion of halo concentration as a function of mass and
redshift are explained by the spread in halo mass accretion
histories.}

\section{Halo Centers}

Already in the early 1990s, high resolution simulations of individual
galaxy halos in CDM were finding $\rho(r) \sim r^{-\alpha}$ with
$\alpha \sim 1$.  This behavior implies that the rotation velocity at
the centers of galaxies should increase as $r^{1/2}$, but the data,
especially that on dark-matter-dominated dwarf galaxies, instead
showed a linear increase with radius, corresponding to roughly
constant density in the centers of galaxies.  This disagreement of
theory with data led to concern that CDM might be in serious
trouble \cite{florespri94,moore94}.

Subsequently, NFW \cite{nfw} found that halos in all variants of CDM
are well fit by the $\rho_{NFW}(r)$ given above, while Moore's group
proposed an alternative $\rho_M(r) \propto r^{-3/2}(r+r_M)^{-3/2}$
based on a small number of very-high-resolution simulations of
individual halos \cite{moore98,moore99,moore00}.  Klypin and
collaborators (including me) initially claimed that typical CDM halos
have shallow inner profiles with $\alpha \approx 0.2$ \cite{klyp98},
but we subsequently realized that the convergence tests that we had
performed on these simulations were inadequate.  We now have simulated
a small number of galaxy-size halos with very high resolution
\cite{klypin01}, and find that they range between $\rho_{NFW}$ and
$\rho_M$.  Actually, these two analytic density profiles are
almost indistinguishable unless galaxies are probed at scales
smaller than about 1 kpc.

Meanwhile, the observational situation is improving.  The rotation
curves of dark matter dominated low surface brightness (LSB) galaxies
were measured with radio telescopes during the 1990s, and the rotation
velocity was typically found to rise linearly at their centers
\cite{bmh,mcgaugh,klyp98}. But a group led by van den Bosch
\cite{vdbetal} showed that in many cases the large beam size of the
radio telescopes did not adequately resolve the inner parts of the
rotation curves, and they concluded that after correcting for beam
smearing the data are on the whole consistent with expectations from
CDM.  Similar conclusions were reached for dwarf galaxies
\cite{vdbswaters}. Swaters and collaborators showed that optical
(H$\alpha$) rotation curves of some of the LSB galaxies rose
significantly faster than the radio (HI) data on these same galaxies
\cite{swatersmt}, and these rotation curves (except for
F568-3) appear to be more consistent with NFW \cite{swatersrome}. 

Recently, a large set of high-resolution optical rotation curves
has been analyzed for LSB galaxies, including many new observations
\cite{deblok01}. The first conclusion that I reach in looking at the
density profiles presented is that the NFW profile often appears to be
a good fit down to about 1 kpc.  However, some of these galaxies
appear to have shallower density profiles at smaller radii.  Of the 48
cases presented (representing 47 galaxies, since two different data
sets are shown for F568-3), in a quarter of the cases the data do not
probe inside 1 kpc, and in many of the remaining cases the resolution
is not really adequate for definite conclusions, or the interpretation
is complicated by the fact that the galaxies are nearly edge-on.  Of
the dozen cases where the inner profile is adequately probed, about
half appear to be roughly consistent with the cuspy NFW profile (with
fit $\alpha \gsim 0.5$), while half are shallower.  This is not
necessarily inconsistent with CDM, since observational biases such as
seeing and slight misalignment of the slit lead to shallower profiles
\cite{swatersVenice}.  Perhaps it is significant that the cases where
the innermost data points have the smallest errors are cuspier.

I think that this data set may be consistent with an inner density
profile $\alpha \sim 1$ but probably not steeper, so it is definitely
inconsistent with the claims of the Moore group that $\alpha \gsim
1.5$.  But recent work~\cite{Power} has shown that Moore's simulations did
not have adequate resolution to support their claimed steep central
cusp; the highest-resolution simulations appear to be consistent with
NFW, or even shallower with $\alpha \approx 0.75$.  Further
simulations and observations, including measurement of CO rotation
curves~\cite{blitz}, may help to clarify the nature of the dark
matter.

It is something of a scandal that, after all these years of simulating
dark matter halos, we still do not have a quantative --- or even a
qualitative --- theory explaining their radial density profiles.  In
her dissertation research \cite{risa01}, Risa Wechsler found that the
central density profile and the value of $r_s$ are typically
established during the early, rapidly merging phase of halo evolution,
and that, during the usually slower mass accretion afterward, $r_s$
changes little (see also \cite{zhao}).  The mass added on the halo
periphery increases $R_{vir}$, and thus the concentration $c_{vir}
\equiv R_{vir}/r_s$.  Now we want to understand this analytically.
Earlier attempts to model the result of sequences of mergers (e.g.,
\cite{syerwhite,nussersheth}) led to density profiles that depend
strongly on the power spectrum of initial fluctuations, in conflict
with simulations (e.g.~\cite{hussjs}).  Perhaps it will be possible to
improve on the simple analytic model of mass loss due to tidal
stripping during satellite inspiral that we presented in
\cite{bullock00}.  Including the tidal puffing up of the inspiralling
satellite before tidal stripping can perhaps account for the origin of
the cusp seen in dissipationless simulations, independent of the power
spectrum~\cite{dekeldevor}.  They argue that the profile must be
steeper than $\alpha=1$ as long as enough satellites make it into the
halo inner regions, simply because for flatter profiles the tidal
force causes dilation rather than stripping.  The proper modeling of
the puffing and stripping in the merger process of CDM halos may also
provide a theoretical framework for understanding the observed flat
cores as a result of gas processes; reionization and feedback into the
baryonic component of small satellites would make their cores puff up
before merging. This could cause them to be torn apart before they
penetrate into the halo centers, and thus allow $\alpha <1$
cores \cite{mallerdekel}. Other possible explanations for flatter
central density profiles involving the baryonic component in galaxies
has recently been proposed, in which the baryons form a bar that
transfers angular momentum into the inner parts of the
halo\cite{katzweinberg}, or alternatively binary black holes eject
matter by a gravitational slingshot effect~\cite{Merritt}. While these
phenomena could be very important in massive galaxies, it is not clear
that they are important in dark-matter-dominated dwarf and LSB
galaxies that have small or nonexistent bulge components.

It would be interesting to see whether CDM can give a consistent
account of the distribution of matter near the centers of big
galaxies, but this is not easy to test.  One might think that big
bright galaxies like the Milky Way could help to test the predicted
CDM profile, but the centers of such galaxies are dominated by
ordinary matter (stars) rather than dark matter.\footnote{Navarro and
Steinmetz had claimed that the Milky Way is inconsistent with the NFW
profile \protect\cite{navstein00}, but they have now shown that \lcdm\
simulations with a proper fluctuation spectrum are actually consistent
with the data \protect\cite{navstein01}.}

\section{Too Much Substructure?}

Another concern is that there are more dark halos in CDM simulations
with circular velocity $V_c \lsim 30$ km s$^{-1}$ than there are
low-$V_c$ galaxies in the Local Group \cite{klypsat,mooresat}. A
natural solution to this problem was proposed by Bullock et
al. \cite{bkw00}, who pointed out that gas will not be able to cool in
$V_c \lsim 30$ km s$^{-1}$ dark matter halos that collapse after the
epoch of reionization, which occured perhaps at redshift $z_{reion}
\approx 6$~\cite{fan}.  When this is taken into account, the predicted
number of small satellite galaxies in the Local Group is in good
agreement with observations \cite{bkw00,moore01}. It is important to
develop and test this idea further, and this is being done by James
Bullock and by Rachel Somerville and their collaborators; the results
to date (e.g.~\cite{bkw01,squelching}) look rather promising.  Other
groups (e.g.~\cite{scann,chiu,benson}) now agree that astrophysical
effects will keep most of the subhalos dark.  As a result, theories
such as warm dark matter (WDM), which solve the supposed problem of
too many satellites by decreasing the amount of small scale power, may
end up predicting too few satellites when reionization and other
astrophysical effects are taken into account~\cite{barkana}.

The fact that high-resolution CDM simulations of galaxy-mass halos are
full of subhalos has also led to concerns that all this substructure
could prevent the resulting astrophysical objects from looking like
actual galaxies \cite{mooresat}. In particular, it is known that
interaction with massive satellites can thicken or damage the thin
stellar disks that are characteristic of spiral galaxies, after the
disks have formed by dissipative gas processes.  However, detailed
simulations~\cite{walker,velazquezwhite} have shown that simpler
calculations~\cite{tothost} had overestimated the extent to which small
satellites could damage galactic disks.  Only interaction with large
satellites like the Large Magellanic Cloud could do serious damage.
But the number of LMC-size and larger satellites is in good agreement
with the number of predicted halos~\cite{klypsat}, which suggests that
preventing disk damage will not lead to a separate constraint on halo
substructure.

\section{Angular Momentum Problems}

As part of James Bullock's dissertation research, we found that the
distribution of specific angular momentum in dark matter halos has a
universal profile~\cite{bullock00}.  But if the baryons have the same
angular momentum distribution as the dark matter, this implies that
there is too much baryonic material with low angular momentum to form
the observed rotationally supported exponential
disks~\cite{bullock00,vdbosch01}.  It has long been assumed
(e.g.~\cite{bffp,mmw}) that the baryons and dark matter in a halo
start with a similar distribution, based on the idea that angular
momentum arising from large-scale tidal torques will be similar across
the entire halo.  But as my colleagues and I argued recently, a key
implication of our new picture of angular momentum growth by
merging~\cite{vitvitska} is that the DM and baryons will get different
angular momentum distributions.  For example, the lower density gas
will be stripped by pressure and tidal forces from infalling
satellites, and in big mergers the gaseous disks will partly become
tidal tails.  Feedback is also likely to play an important role, and
Maller and Dekel \cite{mallerdekel} have shown using a simple model that
this can account for data on the angular momentum distribution in low
surface brightness galaxies~\cite{vdboschetal01}.

A related concern is that high-resolution hydrodynamical simulations
of galaxy formation lead to disks that are much too small, evidently
because formation of baryonic substructure leads to too much transfer
of angular momentum and energy from the baryons to the dark matter
\cite{navsteindisk}. But if gas cooling is inhibited in the early
universe, more realistic disks form \cite{weil}, more so in
$\Lambda$CDM than in $\Omega_m=1$ CDM~\cite{eke}.  Hydrodynamical
simulations also indicate that this disk angular momentum problem may
be resolved if small scale power is suppressed because the dark matter
is warm rather than cold \cite{sldolgov}, which I discuss next.

\section{Alternatives to $\Lambda$CDM?}  

Because of the concerns just mentioned that CDM may predict higher
densities and more substructure on small scales than is observed, many
people have proposed alternatives to CDM.  Two of these ideas that
have been studied in the greatest detail are self-interacting dark
matter (SIDM) \cite{sidm} and warm dark matter (WDM).

Cold dark matter assumes that the dark matter particles have only weak
interactions with each other and with other particles.  SIDM assumes
that the dark matter particles have strong elastic scattering cross
sections, but negligible annihilation or dissipation.  The hope was
that SIDM might suppress the formation of the dense central regions of
dark matter halos, although the large cross sections might also lead
to high thermal conductivity which drains energy from halo centers and
could lead to core collapse~\cite{burkert}, and which also causes
evaporation of galaxy halos in clusters, resulting in violation of the
observed ``fundamental plane'' correlations~\cite{gnedinost}. But in
any case, self-interaction cross sections large enough to have a
significant effect on the centers of galaxy-mass halos will make the
centers of galaxy clusters more spherical \cite{miralda,yoshida} and
perhaps also less dense \cite{meneghetti,eturner} than gravitational
lensing observations \cite{arabadjis} indicate.

Warm dark matter arises in particle physics theories in which the dark
matter particles have relatively high thermal velocities, for example
because their mass is $\lsim 1$ keV \cite{pp82}, comparable to the
temperature about a year after the Big Bang when the horizon first
encompassed the amount of dark matter in a large galaxy.  Such a
velocity distribution can suppress the formation of structure on small
scales.  Indeed, this leads to constraints on how low the WDM particle
mass can be.  From the requirement that there is enough small-scale
power in the linear power spectrum to reproduce the observed
properties of the Ly$\alpha$ forest in quasar spectra, it follows that
this mass must exceed about 0.75 keV \cite{narayanan}. The requirement
that there be enough small halos to host early galaxies to produce the
floor in metallicity observed in the Ly$\alpha$ forest systems, and
early galaxies and quasars to reionize the universe, probably implies
a stronger lower limit on the WDM mass of at least 1 keV
\cite{haiman}. Simulations \cite{colin00,bode} do show that there will
be far fewer small satellite halos with $\Lambda$WDM than
$\Lambda$CDM.  However, as I have already mentioned, inclusion of the
effects of reionization may make the observed numbers of satellite
galaxies consistent with the predictions of $\Lambda$CDM \cite{bkw00},
in which case $\Lambda$WDM may predict too few small satellite
galaxies \cite{bullocktilt}.  Lensing can be used to look for these
subhalos~\cite{madaumetcalf,chiba} and may already indicate that there
are more of them than expected in $\Lambda$WDM~\cite{kochanek01}.
Thus it appears likely that WDM does not solve all the problems it was
invoked to solve, and may create new problems.  Moreover, even with an
initial power spectrum truncated on small scales, simulations appear
to indicate that dark matter halos nevertheless have density profiles
much like those in CDM \cite{huss,moore99,navstein01} (although doubts
have been expressed about the reliability of such simulations because
of numerical relaxation \cite{dalhogan}).  But WDM does lead to lower
concentration halos in better agreement with observed rotation
velocity curves~\cite{avila,alam}.

One theoretical direction that does appear very much worth
investigating is \lcdm\ with a tilt $n\sim 0.9$ in the primordial
power spectrum $P_p(k) \propto k^n$ \cite{bullocktilt}.  Such t\lcdm\
cosmology is favored by recent measurements of the power spectrum of
the Ly $\alpha$ forest \cite{croft} and appears to be consistent with
the latest CMB measurements and all other available
data~\cite{wang01}.  Our simple analytic model~\cite{bullock99}
predicts that the concentration of halos in t\lcdm\ will be
approximately half that in LCDM, because the reduced power on small
scales makes the halos form later.  While this does not resolve the
possible cusp problem, it is a step in the right direction which may
lessen the conflict with galaxy rotation curves.

\section{Conclusion}

The successes of the CDM paradigm are remarkable.  Except possibly for
the density profiles at the centers of dwarf and low surface
brightness galaxies, the predictions of $\Lambda$CDM appear to be in
good agreement with the available observations.  The disagreements
between predictions and data at galaxy centers appear to occur on
smaller scales than was once thought.  As the data improve it is
possible that the discrepancies on $\lsim 1$ kpc scales may ultimately
show that CDM cannot be the correct theory of structure formation.
However, \lcdm\ appears to be better than any alternative theory that
has so far been studied, even though these alternative theories have
additional adjustable parameters.  Maybe \lcdm\ is even true.

{\def\it{}
\def\bf{}

\end{document}

\end{document}

\end{document}

\bibitem{isss} J.R. Primack, in {\it COSMO-2000, Proc. 4th
International Workshop on Particle Physics and the Early Universe}
(Jeju Island, Korea, September 2000), eds. J.E. Kim et al. (World
Scientific, Singapore, 2001), pp. 1-18, and J.R. Primack, in
Proceedings of International School of Space Science 2001, ed. Aldo
Morselli (Frascati Physics Series), astro-ph/0112255.

\bibitem{quintessence} P.J. Steinhardt  {\it Physica Scripta} 
{\bf 185}, 177 (2000) and references therein.

\bibitem{peeb82} P.J.E. Peebles {\it ApJ} {\bf 263}, L1  (1982).

\bibitem{pb83} J.R. Primack and G.R. Blumenthal in {\it Formation and
Evolution of Galaxies and Large Structures in the Universe}, eds.
J.~Audouze and J.~Tran Thanh Van (Reidel, Dordrecht, 1983), p. 163.

\bibitem{bfpr} G.R. Blumenthal, S.M. Faber, J.R.  Primack, and
M.J. Rees, {\it Nature} {\bf 311}, 517 (1984).

\bibitem{sidm} D.N. Spergel and P.J. Steinhardt,
{\it Phys. Rev. Lett.} {\bf 84}, 3760 (2000).

\bibitem{dm2000} J.R. Primack, in {\it Proc. 4th International
Symposium on Sources and Detection of Dark Matter in the Universe} (DM
2000), ed. D. Cline (Springer, Berlin, 2001), p. 3.

\bibitem{lange} A.E. Lange et al., {\it Phys. Rev.} D{\bf 63}, 042001
(2000); P. de Bernardis et al., {\it ApJ} submitted, astro-ph/0105296 (2001).

\bibitem{maxima} A.  Balbi et al., {\it ApJ}, {\bf 545}, L1 (2000),
erratum {\bf 558}, L145 (2001); R. Stompor et al., {\it ApJ}, {\bf
561}, L7 (2001).

\bibitem{jaffe} A.H. Jaffe et al., astro-ph/0007333  (2000).

\bibitem{bond} J.R. Bond et al., astro-ph/0011378, in {\it Proc. IAU
Symposium 201}, eds. A. Lasenby and A. Wilkinson (Astronomical Society
of the Pacific, San Francisco, 2002).

\bibitem{pryke} C. Pryke et al., {\it ApJ} submitted, astro-ph/0104490
(2001).

\bibitem{SNIa_IR} A.G. Riess et al., {\it ApJ}, {\bf 536}, 62 (2000);
{\it ApJ}, {\bf 560}, 49 (2001).

\bibitem{SNAP} The SNAP proposal can be downloaded from http://snap.lbl.gov.

\bibitem{freedman} W.L. Freedman et al., {\it ApJ} {\bf 553}, 47 (2001).

\bibitem{Carlstrom} J.E. Carlstrom et al., {\it Physica Scripta} T {\bf
85}, 148 (2000); J.E. Carlstrom et al., astro-ph/0103480, to 
appear in {\it Constructing the Universe with Clusters of Galaxies}, 
eds. F. Durret and G. Gerbal.

\bibitem{carretta} E. Carretta, R. Gratton, G. Clementini, and F. Fusi
Pecci, {\it ApJ} {\bf 533}, 215 (2000).

\bibitem{sneden} C. Sneden et al., {\it ApJ} {\bf 536}, L85 (2000); R. 
Cayrel et al., {\it Nature} {\bf 409}, 691 (2001).

\bibitem{Whi} S.D.M. White, G. Efstathiou, C.S. and Frenk, {\it MNRAS}
{\bf 262}, 1023 (1993).

\bibitem{Evr} A.E. Evrard, C.A. Metzler, and J.F. Navarro, {\it ApJ}
{\bf 469}, 494 (1996).

\bibitem{Kir} D. Kirkman, D. Tytler, S. Burles, D. Lubin, and J.M.
O'Meara {\it ApJ} {\bf 529}, 655 (2000).

\bibitem{Tytler00} D. Tytler, J.M. O'Meara, N. Suzuki, and D. Lubin,
{\it Physica Scripta} {\bf T85}, 12 (2000).

\bibitem{Mohr} J.J. Mohr, B. Mathiesen, and A.E. Evrard, {\it ApJ}
{\bf 517}, 627 (1999).

\bibitem{ekeetal} V.R. Eke, S. Cole, C.S. Frenk, and J.P. Henry, 
{\it MNRAS} {\bf 298}, 1145 (1998).

\bibitem{henry} J.P. Henry, {\it ApJ} {\bf 534}, 565 (2000).

\bibitem{Gonzalez} A.H. Gonzalez, D. Zaritsky, J.J. Dalcanton, and
A.E.  Nelson, in {\it Clustering at High Redshift}, ed. A. Mazure et
al., ASP Conference Series, Vol. 200, 416 (2000); A.H. Gonzalez (2000)
UCSC Ph.D. dissertation.

\bibitem{croft} D.H. Weinberg et al., (1999) {\it ApJ} {\bf 522}, 563; but cf. 
R.A. Croft et al., astro-ph/0012324 (2001).

\bibitem{powerspectrum} W.J. Percival et al., (2dF), {\it MNRAS}, {\bf
327}, 1297 (2001); A.S. Szalay et al., (SDSS), astro-ph/0107419 (2001);
cf. (2MASS) B. Allgood, G.R. Blumenthal, and J.R. Primack, in
Proc. Marseille 2001 Conf. {\it Where's the matter? Tracing dark and
and bright matter with the new generation of large scale surveys}, eds
R. Treyer \& L. Tresse, astro-ph/0109403 (2001).

\bibitem{kph} A.A. Klypin, J.R. Primack, and J. Holtzman,  {\it ApJ}
{\bf 466}, 13 (1996).

\bibitem{jenkins} A. Jenkins et al., {\t ApJ} {\bf 499}, 20 (1998).

\bibitem{overcoming} A.A. Klypin, S. Gottl\"{o}ber, A.V. Kravtsov, 
and A.M. Khokhlov {\it ApJ} {\bf 516}, 530 (1999).

\bibitem{colin} P. Colin, A.A. Klypin, A.V. Kravtsov, and
A.M. Khokhlov, {\it ApJ} {\bf 523}, 32 (1999).

\bibitem{Baugh96} C.M. Baugh, {\it MNRAS} {\bf 280}, 267 (1996).

\bibitem{bffp} G.R. Blumenthal, S.M. Faber, R. Flores, and J.R.
Primack, {\it ApJ} {\bf 301}, 27 (1986); R. Flores, J.R. Primack,
G.R. Blumenthal, and S.M. Faber,  {\it ApJ} {\bf 412}, 443 (1993).

\bibitem{mmw} H.-J. Mo, S. Mao, and S.D.M. White,  {\it MNRAS} 
{\bf 295}, 319 (1998).

\bibitem{kochanekwhite} C.S. Kochanek and M. White, {\it ApJ}, {\bf
559}, 531 (2001).

\bibitem{vfun} A.H. Gonzalez, K.A. Williams, J.S. Bullock, T.S.
Kolatt, and J.R. Primack, {\it ApJ} {\bf 528}, 145 (2000).

\bibitem{cole00} S. Cole, C.  Lacey, C. Baugh, and C.S. Frenk, 
 {\it MNRAS} {\bf 319}, 168 (2000).

\bibitem{davispeeb} M. Davis and P.J.E. Peebles, {\it ApJ} {\bf 267},
465 (1983).

\bibitem{2dF} N. Cross, S, Driver, and W. Couch (2dF Collaboration),
astro-ph/0012165, MNRAS in press (2001).

\bibitem{sdss}  M.R. Blanton et al., (SDSS Collaboration), {\it AJ}
{\bf 121}, 2358 (2001).

\bibitem{pahre} M.A. Pahre et al. (MALIGN survey), AAS Meeting 197,
\#96.01 (2000).

\bibitem{Weinberg}  D.H. Weinberg, L. Hernquist, and N. Katz, 
{\it ApJ} submitted astro-ph/0005340 (2000).

\bibitem{whitefrenk}  S.D.M. White and C.S. Frenk,  {\it ApJ}
{\bf 379}, 52 (1991).

\bibitem{kauffmann93} G. Kauffmann,  S.D.M. White, and B. Guiderdoni, 
{\it MNRAS} {\bf 264}, 201 (1993).

\bibitem{cole94} S. Cole et al., {\it MNRAS} {\bf 271}, 781 (1994).

\bibitem{sp} R.S. Somerville and J.R. Primack, {\it MNRAS}, {\bf 310},
1087 (1999).

\bibitem{spf}  R. S. Somerville, J.R. Primack, and S.M. Faber,
{\it MNRAS} {\bf 320}, 504 (2001).

\bibitem{nfw} J.F. Navarro, C.D. Frenk, C.S. and S.D.M. White, {\it
ApJ} {\bf 462}, 563 (1996); {\it ApJ} {\bf 490}, 493 (1997).

\bibitem{klypin01} A.A. Klypin, A.V. Kravtsov, J.S. Bullock, and J.R.
Primack, {\it ApJ} {\bf 554}, 903 (2001).

\bibitem{art}  A.V. Kravtsov, A.A. Klypin, and A.M. Khokhlov, 
{\it ApJS} {\bf 111}, 73 (1997).

\bibitem{bullock99} J.S. Bullock, T..S. Kolatt, Y, Sigad, R,S.
Somerville, A.V. Kravtsov, A.A. Klypin, J.R. Primack, and A.
Dekel, {\it MNRAS} {\bf 321}, 559 (2001).

\bibitem{jing}  Y. Jing, {\it ApJ} {\bf 535}, 30 (2000).

\bibitem{risa01} R.H. Wechsler, J.S. Bullock, J.R. Primack,
A.V. Kravtsov, and A. Dekel, {\it ApJ} in press, astro-ph/0108151
(2001).

\bibitem{vdboschTF}  F.C. van den Bosch (2000) {\it ApJ} {\bf 530}, 177.

\bibitem{florespri94} R.A. Flores and J.R. Primack,  {\it ApJ},
{\bf 427}, L1 (1994).

\bibitem{moore94} B. Moore, {\it Nature} {\bf 370}, 620 (1994).

\bibitem{moore98} B. Moore et al., {\it ApJ} {\bf 499}, L5 (1998).

\bibitem{moore99} B. Moore et al., {\it MNRAS} {\bf 310}, 1147 (1999).

\bibitem{moore00} S. Ghigna et al., {\it ApJ} {\bf 544}, 616 (2000).

\bibitem{bmh} W.J.G. de Blok, S.S. McGaugh, and J.M. van der Hulst,
 {\it MNRAS} {\bf 283}, 18 (1996).

\bibitem{mcgaugh} S.S. McGaugh and W.J.G.  de Blok, {\it ApJ} {\bf
499}, 41 (1998) and references cited therein.

\bibitem{klyp98} A.A. Klypin, A.V. Kravtsov, J.S. Bullock, and J.R.
Primack, {\it ApJ} {\bf 502}, 48 (1998) and references cited therein.

\bibitem{vdbetal} F.C. van den Bosch, B.E. Robertson, J.J. Dalcanton,
and W.J.G. de Blok, {\it AJ}, {\bf 119}, 1579 (2000).

\bibitem{vdbswaters} F. van den Bosch and R.A. Swaters, {\it MNRAS}
{\bf 325}, 1017 (2001).

\bibitem{swatersmt}  R.A. Swaters, B.F. Madore, and M. Trewhella, 
{\it ApJ} {\bf 531}, L107 (2000).

\bibitem{swatersrome} R.A. Swaters, in {\it Galaxy Disks and Disk
Galaxies}, ed. J. G. Funes and E.M. Corsini, ASP Conf. Series,
Vol. 230, p. 545 (2001).

\bibitem{deblok01} W.J.G. de Blok, S.S. McGaugh, A. Bosma, and V.C.
Rubin, {\it ApJ}, 552, L23 (2001);  S.S. McGaugh, V.C. Rubin, and
W.J.G. de Blok, {\it AJ} {\bf 122}, 2381 (2001);  W.J.G. de Blok,
S.S. McGaugh, and V.C. Rubin, {\it AJ} {\bf 122}, 2396 (2001).

\bibitem{swatersVenice} R. Swaters, in Proc. {\it The Mass of Galaxies
at Low and High Redshift}, Venice Oct 2001, eds. R. Bender and
A. Renzini (Springer-Verlag, 2002).

\bibitem{navarro01} J.F. Navarro, in IAU Symposium 208, {\it
Astrophysical SuperComputing using Particles}, eds. J. Makino and P.Hut,
astro-ph/0110680 (2001).

\bibitem{blitz} A.D. Bolatto, J.D. Simon, A. Leroy, and L. Blitz, {\it
ApJ} in press astro-ph/0108050 (2001).

\bibitem{syerwhite} D. Syer and S.D.M. White, {\it MNRAS} {\bf
293}, 377 (1998).

\bibitem{nussersheth} A. Nusser and R.K. Sheth, {\it MNRAS} {\bf 303},
685 (1999).

\bibitem{hussjs} A. Huss, B. Jain, and M. Steinmetz, {\it ApJ} {\bf
517}, 64 (1999).

\bibitem{katzweinberg} M.D. Weinberg and N. Katz, astro-ph/0110632
(2001).

\bibitem{navstein00} J.F. Navarro and M. Steinmetz, {\it ApJ} {\bf
528}, 607 (2000).

\bibitem{navstein01} V.R. Eke, J.F. Navarro, and M. Steinmetz,
{\it ApJ} {\bf 554}, 114 (2001).

\bibitem{klypsat} A.A. Klypin, A.V. Kravtsov, O. Valenzuela, and
F. Prada, {\it ApJ} {\bf 522}, 82 (1999).

\bibitem{mooresat} B. Moore et al., ApJ 524, L19 (1999).

\bibitem{bkw00} J.S. Bullock, A.V.  Kravtsov, and D.H. Weinberg, 
 {\it ApJ} {\bf 539}, 517 (2000).

\bibitem{fan} See X. Fan et al., {\it AJ} submitted, astro-ph/0111184
(2001), and references therein.

\bibitem{moore01} B. Moore, astro-ph/0103100, in {\it 20th Texas
Symposium}, eds. J. C. Wheeler and H. Martel, in press (2001).

\bibitem{bkw01} J.S. Bullock, A.V. Kravtsov, and D.H. Weinberg, {\it
ApJ} {\bf 548}, 33 (2001).

\bibitem{squelching} R.S. Somerville, astro-ph/0107507 (2001).

\bibitem{scann} E. Scannapieco, R.J. Thacker, and M. Davis,
{\it ApJ} {\bf 557}, 605 (2001).

\bibitem{chiu} W.A. Chiu, N.Y. Gnedin, and J.P. Ostriker, 
astro-ph/0103359 (2001).

\bibitem{benson} A.J. Benson et al., astro-ph/0108218 (2001).

\bibitem{barkana} R. Barkana, Z. Haiman, and J.P. Ostriker, 
{\it ApJ} {\bf 558}, 482 (2001).

\bibitem{walker} I.R. Walker, J.C. Mihos, and L. Hernquist, 
{\it ApJ} {\bf 460}, 121 (1996).

\bibitem{velazquezwhite} H. Velazquez and S.D.M. White,
{\it MNRAS} {\bf 304}, 254 (1999).

\bibitem{tothost} G. Toth and J.P. Ostriker, {\it ApJ} {\bf 389}, 5
(1992).

\bibitem{bullock00} J.S. Bullock, A. Dekel, T.S. Kolatt,
A.V. Kravtsov, C. Porciani, and J.R. Primack,
{\it ApJ} {\bf 555}, 240 (2000).

\bibitem{vdbosch01} F. van den Bosch, {\it MNRAS} {\bf 327}, 1334 (2001).

\bibitem{vitvitska} M. Vitvitska, A.A. Klypin, A.V. Kravtsov, J.S.
Bullock, R.H. Wechsler, and J.R. Primack, {\it ApJ} submitted,
astro-ph/0105349; A.H. Maller, A. Dekel, and R.S. Somerville,
{\it MNRAS} submitted, astro-ph/0105168 (2001).

\bibitem{vdboschetal01} F. van den Bosch, A. Burkert, and
R.A. Swaters, {\it MNRAS} {\bf 326}, 1205 (2001).

\bibitem{navsteindisk} J.F. Navarro and M. Steinmetz, {\it ApJ} {\bf
478}, 13 (1997); {\bf 513}, 555 (1999); {\bf 538}, 477 (2000).

\bibitem{weil}  M.L. Weil,  V.R. Eke, and g. Efstathiou,  {\it MNRAS} 
{\bf 300}, 773 (1998).

\bibitem{eke} V.R. Eke, G. Efstathiou, and L. Wright, {\it MNRAS} {\bf
315}, L18 (2000).

\bibitem{sldolgov} J. Sommer-Larsen and A. Dolgov, {\it ApJ} {\bf
551}, 608 (2001).

\bibitem{burkert} A. Burkert, astro-ph/0012178 (2000).

\bibitem{gnedinost} O.Y. Gnedin and J.P. Ostriker, {\it ApJ} {\bf
561}, 61 (2001).

\bibitem{miralda} J. Miralda-Escude, astro-ph/0002050 (2000).

\bibitem{yoshida} N. Yoshida, V. Springel, S.D.M.  White, and
G. Tormen, {\it ApJ} {\bf 544}, L87 (2000).

\bibitem{meneghetti} M. Meneghetti et al., {\it MNRAS} {\bf 325}, 435 (2001).

\bibitem{eturner}  J.S.B. Wyithe, E.L. Turner, and D.N. Spergel,
{\it ApJ} {\bf 555}, 504 (2001).

\bibitem{arabadjis} See e.g.~J.S. Arabadjis, M.W. Bautz, and
G.P. Garmire, astro-ph/0109141 (2001).

\bibitem{pp82} H. Pagels and J.R. Primack, {\it Phys. Rev. Lett.}
{\bf 48}, 223 (1982); G.R. Blumenthal, H. Pagels, and J.R. Primack,
{\it Nature} {\bf 299}, 37 (1982).

\bibitem{narayanan} V.K. Narayanan, D.N. Spergel, R. Dav\'e, and
C.P. Ma, {\it ApJ} {\bf 543}, L103 (2000).

\bibitem{haiman} Z. Haiman, R. Barkana, and J.P. Ostriker, 
astro-ph/0103050 (2001).

\bibitem{colin00} R. Colin, V. Avila-Reese, and O. Valenzuela, 
{\it ApJ} {\bf  542}, 622 (2000).

\bibitem{bode} P. Bode, J.P. Ostriker, and N. Turok, {\it ApJ}, {\bf
556}, 93 (2001).

\bibitem{madaumetcalf} R.B. Metcalf and P. Madau, {\it ApJ} in press,
astro-ph/0108224 (2001).

\bibitem{chiba} M. Chiba, {\it ApJ} in press,  astro-ph/0109499 (2001).

\bibitem{kochanek01} N. Dalal and C.H. Kochanek, astro-ph/0111456 (2001).

\bibitem{huss} A. Huss, B. Jain, and M. Steinmetz, {\it ApJ} {\bf
517}, 64 (1999).

\bibitem{dalhogan} J.J. Dalcanton and C.J. Hogan, {\it ApJ} {\bf 561},
35 (2001).

\bibitem{avila} V. Avila-Rees, P. Colin, O. Valenzuela, E. D'Onghia,
and C. Firmani, {\it ApJ} {\bf 559}, 516 (2001).

\bibitem{alam} S.M.K. Alam,  J.S. Bullock,  and D.H. Weinberg, 
{\it ApJ} submitted, astro-ph/0109392 (2001).

\bibitem{bullocktilt} J.S. Bullock, in Proc. Marseille 2001 Conf. {\it
Where's the matter? Tracing dark and and bright matter with the new
generation of large scale surveys}, eds R. Treyer \& L. Tresse,
astro-ph/0111005 (2001).

\bibitem{wang01} X. Wang, M. Tegmark, and M. Zaldarriaga,
astro-ph/0105091 (2001).

\end{thebibliography}
}
\end{document}